\documentclass[apl,reprint,superscriptaddress]{revtex4-1}
\usepackage[english]{babel}
\usepackage{graphicx}
\usepackage{xspace}
\usepackage{fancybox}
\usepackage{epic}
\usepackage{calc}
\usepackage{ifthen}
\usepackage{float}
\usepackage{color}
\usepackage{amsmath,amssymb}
\usepackage{rotating}
\usepackage{subfigure}
\usepackage{exscale}
\usepackage{setspace}
\usepackage[latin1]{inputenc}
\begin{document}
\title{Controlled oxygen vacancy induced $p$-type conductivity in HfO$_{2-x}$ thin films}
\author{Erwin Hildebrandt}
\affiliation{Institute of Materials Science, Technische
Universit\"{a}t Darmstadt, 64287 Darmstadt, Germany}
\author{Jose Kurian}
\affiliation{Institute of Materials Science, Technische
Universit\"{a}t, 64287 Darmstadt, Germany}
\author{Mathis M. Müller}
\affiliation{Institute of Materials Science, Technische
Universit\"{a}t, 64287 Darmstadt, Germany}
\author{Thomas Schroeder}
\affiliation{IHP, 15236 Frankfurt / Oder, Germany}
\author{Hans-Joachim Kleebe}
\affiliation{Institute of Materials Science, Technische
Universit\"{a}t, 64287 Darmstadt, Germany}
\author{Lambert Alff}\email[Author to whom correspondence should be addressed. Electronic mail:\,]{alff@oxide.tu-darmstadt.de}
\affiliation{Institute of Materials Science, Technische
Universit\"{a}t, 64287 Darmstadt, Germany}

\date{09 July 2011}

\begin{abstract}

We have synthesized highly oxygen deficient HfO$_{2-x}$ thin films by controlled oxygen engineering using reactive molecular beam epitaxy. Above a threshold value of oxygen vacancies, $p$-type conductivity sets in with up to 6 times $10^{21}$ charge carriers per cm$^3$. At the same time, the band-gap is reduced continuously by more than 1\,eV. We suggest an oxygen vacancy induced $p$-type defect band as origin of the observed behavior.

\end{abstract}

\maketitle


Hafnia, HfO$_2$, is an insulating material which currently plays an important role in state-of-the-art transistors as replacement of SiO$_2$, the universal gate dielectric, because of its large permittivity \cite{Wilk:01}. Even more, there have been observations of unexpected ferromagnetism with high Curie temperature in hafnia - a behavior which was attributed to oxygen vacancies \cite{Venkatesan:04,Coey:05}. These magnetic properties which could turn HfO$_{2-x}$ also into a favourite material for spintronics, however, are still under debate.
Here, we report on another intriguing observation in hafnium dioxide thin films grown by reactive molecular beam epitaxy (RMBE): Engineering the oxygen vacancies can metamorphose the colorless insulating thin films in a controlled way into a stable highly hole doped semiconductor. In oxides, charge carrier doping can be achieved by the removal of oxygen ions. One major advantage of oxygen engineering is the absence of additional impurity atoms \cite{Mannhart:10} which often implies further constraints like chemical compatibility and the associated undesired consequences in multilayer structures. In some oxides such as SrTiO$_3$ the creation of oxygen vacancies is straightforward, however, introducing oxygen vacancies in a desired amount and controlled manner in an extremely stable oxide like HfO$_2$ remains challenging. Annealing SrTiO$_3$ single crystals or thin films in vacuum at moderately high temperatures induces oxygen vacancies and makes them conducting. The same procedure, in contrast, does not induce oxygen vacancies in an extremely stable oxide like hafnia. Here we demonstrate that by reactive molecular beam epitaxy (RMBE) it is possible to fabricate HfO$_{2-x}$ thin films in which the amount of oxygen vacancies can be varied to such a level that the electrical conductivity can be manipulated within several orders of magnitude. As HfO$_2$ is a material which now plays an important role as high-$k$ dielectric in the Si-based semiconductor devices where the handling and compatibility issues with the Si-based industrial processing are solved, more available functionalities have a potentially high significance.


We have used RMBE in a custom designed, ultra high vacuum chamber with a base pressure of about 10$^{-9}$\,mbar. Hf metal (99.9\% purity, MaTecK) is evaporated by means of electron beam evaporation at a rate between 0.5 and 1.4\,{\AA}/s. We perform in situ oxidation by introducing rf-activated oxygen (99.995\% purity) from an oxygen radical source (power ranging between 0 and 300\,W). The oxygen flow rate used for running the radical source is varied between 0 and 2.5\,sccm. The thin films are deposited on $c$-cut sapphire single crystal substrates (CrysTec). Throughout this study, we have used a typical substrate temperature of 700\,$^\circ$C and film thicknesses of about 100 nm. Hf metal flux control is maintained by a quartz crystal microbalance (QCM) with a control feedback loop. In the chosen growth regime, monoclinic HfO$_{2-x}$ grows in the [111] direction on $c$-cut sapphire. Previously, it was shown that at different temperatures, the growth direction can be changed to [100] \cite{Hildebrandt:09}. We do not observe an oxygen vacancy driven tendency towards the formation of cubic or tetragonal HfO$_{2-x}$ as reported elsewhere \cite{Sarkinos:10}. Using a beam of oxygen radicals in MBE, oxide thin film growth can be achieved even under high vacuum conditions. Under such conditions, non-equilibrium vacancy concentrations can be stabilized and frozen. So far, the oxygen content in hafnium metal oxide thin films has been manipulated by partial pressure adjustment during growth by pulsed laser deposition (PLD) \cite{Coey:05} and by  post deposition processes such as annealing \cite{HWang:08} or ozone treatment \cite{Wang:08}. In comparison, oxide MBE offers a much more precise rate control of Hf and atomic oxygen flux, while at the same time particle energies are thermalized and growth rates slow - allowing for controlled oxygen engineering.

In order to estimate the stoichiometry of our thin films, we have used X-ray photoelectron spectroscopy (XPS). Because in air always a surface layer of fully oxidized HfO$_2$ forms, ion beam cleaning of the samples becomes necessary which is, however, associated with oxygen preferential sputtering leading to unrealistic and sputter time dependent values of oxygen deficiencies in the range of HfO$_{1.3}$. A more meaningful estimation of oxygen defects, charge carrier sign and concentration is obtained by Hall effect measurements. First of all, HfO$_{2-x}$ turns out to have hole-type charge carriers in agreement with a previous observation \cite{Hadacek:07}. In strongly reduced samples, we have measured a charge carrier concentration of about 6 times 10$^{21}$ per cm$^3$. Assuming that one electron is trapped by each oxygen vacancy, this corresponds to about 10\% of active oxygen vacancy sites. The mobility is of the order 2 cm$^2$/(Vs). The higher charge carrier concentration as compared to other $p$-type semiconductors such as (Ga,Mn)As or (Zn,Mn)Te has its prize in a low mobility most likely due to the large amount of defects and also due to the fact that holes in the oxygen $2p$ band generally show low mobility.


\begin{figure}[t]
\centering{%
\includegraphics[width=0.85\columnwidth,clip=]{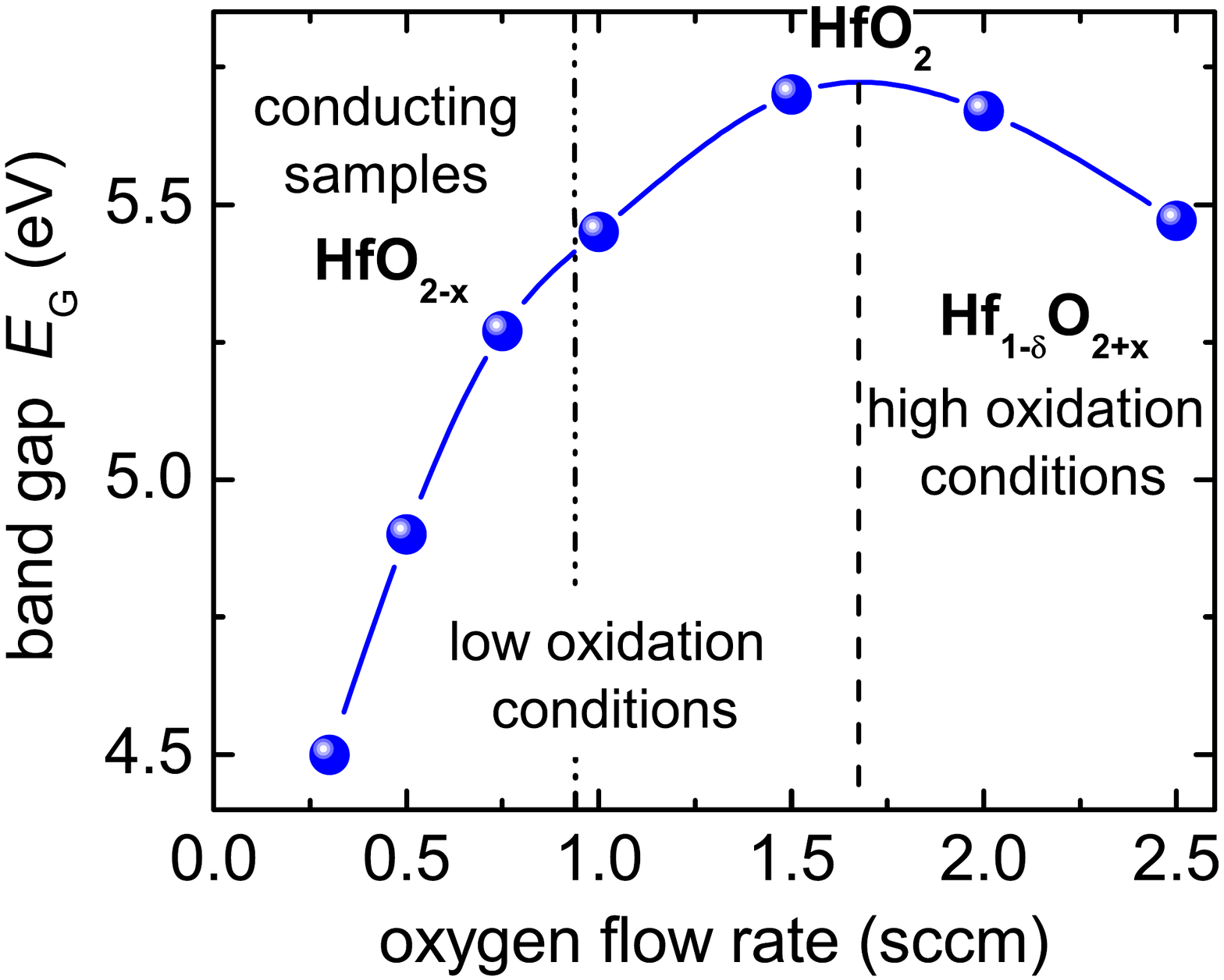}}
\caption{Energy gap of oxygen engineered HfO$_{2-x}$ derived from optical spectroscopy as a function of oxygen flow rate. The maximum corresponds to $x = 0$, the positive slope part corresponds to reduced HfO$_{2-x}$ with $x > 0$ (low oxidation conditions). The negative slope part is associated with the formation of Hf vacancies and oxygen interstitials (high oxidation conditions). The resistivity of the heavily reduced samples left of the dash-dotted line is shown in Fig.~\ref{Fig:resistivity}.}\label{Fig:gap}
\end{figure}

\begin{figure}[t]
\centering{%
\includegraphics[width=0.85\columnwidth,clip=]{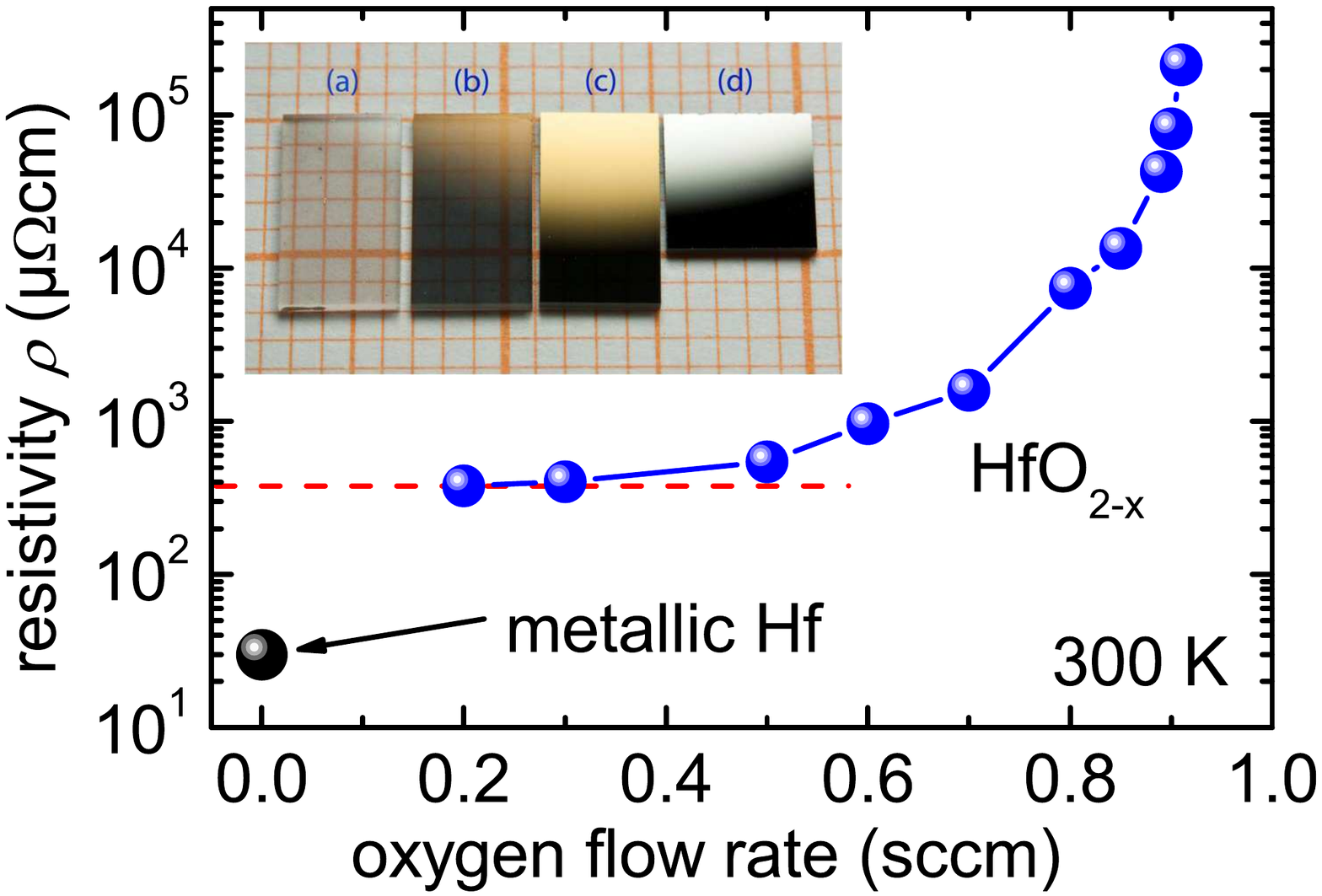}}
\caption{Room-temperature resistivity of HfO$_{2-x}$ as a function of oxygen flow rate.
Inset: Photo of the reflection of a white light bulb from the surface of equally thin films of HfO$_{2-x}$ on $c$-cut sapphire substrates. (a) transparent HfO$_2$ film (2.0\,sccm O$_2$), (b) greyish HfO$_{2-x}$ film (1.0\,sccm), (c) black HfO$_{2-x}$ film, golden shine (0.3\,sccm) and (d) Hf metal thin film, silvery, metallic shine (no oxygen flow).
}\label{Fig:resistivity}
\end{figure}

A first indication of a dramatic change in the materials properties of HfO$_{2-x}$ catches directly the eye (see inset of Fig.~\ref{Fig:resistivity}: Instead of almost perfectly transparent films as obtained for high oxidation conditions, we observe greyish to black thin films grown under low oxygen flow. The golden reflection of white light indicates that the material absorbs photons with energies above 2\,eV, indicating the existence of a subband within the valence band to conduction band energy gap.

We first report the investigation of the energy gap in oxygen engineered hafnium dioxide thin films by optical absorption photospectrometry (the method has been described in \cite{Hildebrandt:09}). The derived band gap clearly correlates with the oxidation conditions (see Fig.~\ref{Fig:gap}). The maximal measured band gap of about 5.7\,eV corresponds to the stoichiometric compound HfO$_2$ \cite{Moon:05}. With decreasing oxygen flow rate, corresponding to increased $x$ in HfO$_{2-x}$, the energy gap is strongly reduced by more than 1\,eV to about 4.5\,eV. With increasing oxygen flow rate beyond the optimal value, the optical band gap also decreases. Both effects can be understood by the point defect chemistry in monoclinic Hf$_{1\pm\delta}$O$_{2\pm x}$ \cite{Tang:10} and the way defect states affect the band structure. In HfO$_2$, energetically oxygen and hafnium vacancies have the lowest formation energies. However, taking into account the kinetics of the growth process, interstitial oxygen defects may also occur under high oxidation conditions because of their reduced migration barrier. The described defects form electronic defect states in the middle of the band gap, but also states which are located in the close vicinity of the conduction band of HfO$_2$ \cite{Pemmaraju:05}. The hybridization of these defect states with the conduction band results in a reduction of the observed optical band gap. These induced defect states and their imprint on the band structure are also important to understand the resistive switching effect in observed for HfO$_2$ \cite{Walczyk:09}


The room temperature resistivity which in insulating, stoichiometric HfO$_2$ is in the range of 10$^{18}\,\mu\Omega$cm, drops steeply as the oxygen flow rate crosses a threshold value. Subsequently, the resistivity spans another three orders of magnitude and runs into a saturation value of about 400\,$\mu\Omega$cm (for comparison: metallic hafnium has a room temperature resistivity of about 35\,$\mu\Omega$cm). The observation that the resistivity of HfO$_{2-x}$ extrapolates to a value of 400\,$\mu\Omega$cm indicates also that we are not dealing with a conducting percolation path via overlapping metallic hafnium clusters inside the sample, but indeed with a defect rich hafnium dioxide.
As a further test, we have annealed insulating HfO$_2$ thin films in a vacuum furnace at 1000\,$^\circ$C for 12\,h. Neither the film nor the Al$_2$O$_3$ substrate showed any traces of conductivity after annealing. This is in strong contrast to the case of SrTiO$_3$ where a vacuum annealing at a moderately higher temperature easily introduces oxygen vacancies within the bulk of the crystal or film. In order to assess the stability of the heavily oxygen vacant material, we have re-measured samples several months after synthesis, and obtained identical resistivity values. These experiments show that it is possible to freeze such high defect concentrations at room temperature on a meaningful time scale.


\begin{figure}[t]
\centering{%
\includegraphics[width=0.85\columnwidth,clip=]{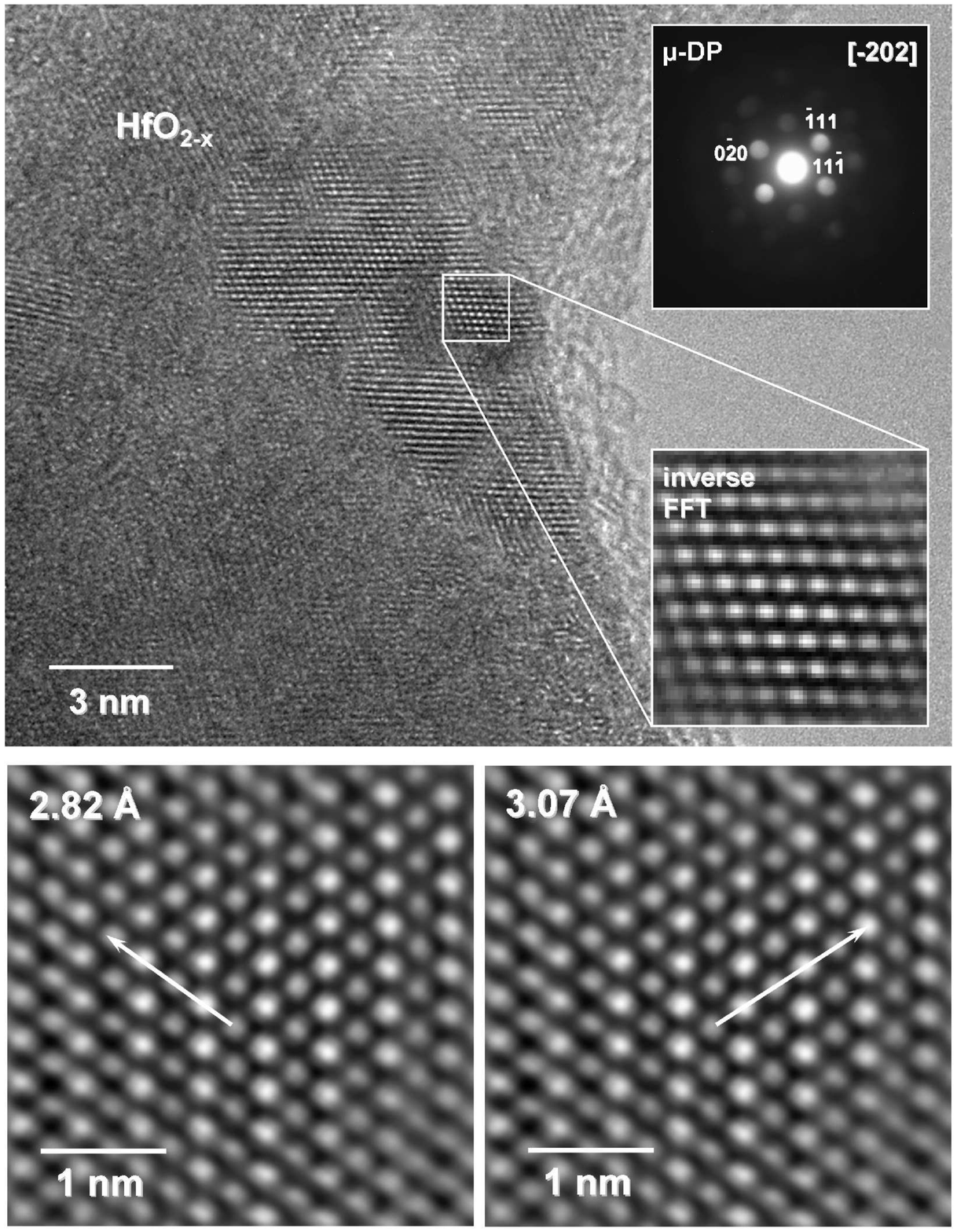}}
\caption{HRTEM images of a conducting HfO$_{2-x}$ thin film. The upper inset reveals a micro-diffraction pattern confirming (i) the textured nature of the thin film and (ii) the formation of HfO$_{2-x}$ in contrast to metallic Hf. The enlarged inverse Fourier-transformed images shown in the lower image represent the lattice periodicity of one individual grain.}\label{Fig:TEM}
\end{figure}

It has been calculated by several authors \cite{Ramo:07,Broqvist:06} that oxygen vacancies in monoclinic HfO$_{2-x}$ can be one or twofold positively charged, neutral, and even one or twofold negatively charged. All the calculations regarding oxygen vacancies in HfO$_{2-x}$ have been performed under the assumption of a low density of oxygen defects. In this case, the energy gap between the valence and the conduction band does not change notably. As observed here, for the case of high defect concentrations, this band gap is, however, substantially reduced from 5.7\,eV to 4.5\,eV. Several defect bands may form due to the overlapping defect states, one of them responsible for the observed $p$-type conductivity. Assuming predominantly single negatively charged defects states overlapping to form a band, $p$-type conductivity may arise according to the curvature of the band at the Fermi level. A mid-gap defect band about 2\,eV away from the conduction band allows for optical transitions in that energy region leading to the observed golden shine (see Fig.~\ref{Fig:resistivity}).

Although $p$-type conductivity excludes the possibility of simple metallic conduction in our samples, an alternative explanation of our data is the presence of metallic Hf clusters that form a percolated conducting path within an insulating or poorly conducting HfO$_2$ matrix. While crystalline conducting phases can be excluded from X-ray diffraction, an amorphous metallic phase can only be excluded by high-resolution transmission electron microscopy (HRTEM). HRTEM (FEI CM20) imaging confirmed (see Fig.~\ref{Fig:TEM}) that our thin films consist of highly textured grains, as verified by electron micro-diffraction. To address as to whether the film also contains an amorphous residue, a tilting series was performed verifying that the HfO$_{2-x}$ film shown in Fig.~\ref{Fig:TEM} was fully crystalline with no amorphous secondary phase present. The analysis of the lattice periodicity, as illustrated in the enlarged inverse Fourier-filtered images, clearly demonstrates the formation of hafnia in contrast to metallic Hf having a hexagonal closed packed structure. The atomic spacings of 0.282 and 0.307\,nm only occur in hafnia and are not present in hafnium metal. It should be noted that those distances are slightly lower than in stoichiometric hafnia (0.282 vs. 0.283 and 0.307 vs. 0.314 nm), which we correlate with the stabilization of oxygen vacancies, proving that we are indeed dealing with defect rich HfO$_{2-x}$. We have not observed any magnetization in our samples as measured by superconducting quantum interference device (SQUID) magnetometry indicating that the presence of charge carriers in HfO$_{2-x}$ is not enough to induce magnetism.


By oxygen engineering using RMBE, we have manipulated electrical transport properties of hafnia in a wide range. The defect states associated with oxygen vacancies seem to form a hole conducting defect band at the Fermi level. At the same time, the band gap between the valence and the conduction band is reduced by about 25\%. The defect chemistry of strongly reduced hafnia still has to be explored theoretically. Taking the high dielectric permittivity, possible ferromagnetism, and now $p$-type semiconducting behavior, HfO$_{2-x}$ is a truely multifunctional material.


This work is supported by the LOEWE Centre of Excellence AdRIA.

\end{document}